\documentclass[runningheads]{llncs}
\usepackage[T1]{fontenc}
\usepackage{graphicx,verbatim,multirow,svg,booktabs,xcolor }
\begin{document}
\raggedbottom 
\sloppy

\title{MRGAgents: A Multi-Agent Framework for Improved Medical Report Generation with Med-LVLMs }
%

\author{Pengyu Wang\inst{1} 
\and Shuchang Ye\inst{1} 
\and Usman Naseem\inst{2} 
\and Jinman Kim\inst{1}}  
\authorrunning{Pengyu Wang. et al.}
\institute{University of Sydney \\
\email{jinman.kim@sydney.edu.au}\\
Macquarie University \\
\email{usman.naseem@mq.edu.au}}
\maketitle              
\begin{abstract}

Medical Large Vision-Language Models (Med-LVLMs) have been widely adopted for medical report generation. Despite Med-LVLMs producing state-of-the-art performance, they exhibit a bias toward predicting findings as normal, leading to reports that overlook critical abnormalities. Furthermore, these models often fail to provide comprehensive descriptions of radiologically relevant regions necessary for accurate diagnosis. To address these challenges, we propose \textbf{M}edical \textbf{R}eport \textbf{G}eneration \textbf{Agents} (\textbf{MRGAgents}), a novel multi-agent framework that fine-tunes specialized agents for different disease categories. By curating subsets of the IU X-ray and MIMIC-CXR datasets to train disease-specific agents, our MRGAgents generates reports that effectively balance normal and abnormal findings while ensuring a comprehensive description of clinically relevant regions. Our experiments demonstrate that MRGAgents outperformed the state-of-the-art, improving both report comprehensiveness and diagnostic utility.

\end{abstract}
\section{Introduction}

Medical imaging plays a critical role in clinical diagnostics, with radiology reports serving as essential tools for communicating diagnostic findings and guiding treatment decisions.  However, the process of generating detailed and accurate medical reports remains time-consuming and labor-intensive \cite{soleimani2024practical,young2025hands}. Automated medical report generation has emerged as a promising solution, offering the potential to reduce radiologists' workload, improve efficiency, and minimize human errors, and thus allowing clinicians to focus more on complex cases and patient care. Over the past few years, significant research efforts have been dedicated to medical report generation (MRG), with models progressively advancing towards more sophisticated  Transformer-based frameworks. These models aim to map complex visual data into coherent textual descriptions, improving the integration of multi-modal information. 

Recent advancements in Large Vision-Language Models (LVLMs) have accelerated the progress in the field, enabling models to process and understand both medical images and textual data more effectively. Notable examples, such as BioMedGPT \cite{zhang2023biomedgpt} and LLaVA-Med \cite{li2023llavamed}, have demonstrated their versatility across various  medical downstream tasks, particularly in MRG \cite{chen-acl-2021-r2gencmn,chen2020generating}. Despite this advancements, existing methods still face significant challenges. First, prior studies \cite{guo2024prompting} have shown that these models exhibit a bias toward normal findings in medical images due to the varying distribution of disease specific description across reports. As a result, wthin a single batch, the model may not adequately learn representation for all disease categories, often overlooking critical abnormalities. This bias results in less informative reports, reducing their clinical utility. Second, reports generated by Med-LVLMs often lack comprehensiveness, which we defined as the ability of a generated medical report to cover all clinically relevant findings, including both normal and abnormal observations, without omitting key diagnostic information.


Meanwhile, recent research demonstrated success of multi-agent collaboration \cite{xi2025rise}, demonstrating its ability to enhance task efficiency, improve decision-making \cite{kim2024mdagents}, and optimize resource allocation across various domains, including healthcare \cite{shi2024ehragent}, autonomous systems, and large-scale data processing. By leveraging the principle of task specialization, multi-agent systems can decompose complex problems into sub-tasks, allowing specialized agents to focus on distinct aspects of the task, thereby improving both efficiency and accuracy \cite{kim2024mdagents,tang2023medagents}. However, existing medical multi-agent frameworks primarily focus on Med-QA (Medical Question Answering) \cite{tang2023medagents} and Med-VQA(Medical Visual Question Answering) \cite{kim2024mdagents,li2024mmedagent,hong2024argmed}, their application to medical report generation remains unexplored. Given that MRG is inherently a complex task requiring fine-gained descriptions across multi-disease categories, we hypothesize that a multi-agent framework can decompose the report generation process into more manageable sub-problems.

In this study, we introduce \textbf{M}edical \textbf{R}eport \textbf{G}eneration \textbf{Agents} (\textbf{MRGAgents}), the\textbf{ first multi-agent LLM-based framework for medical report generation} — that leverages the advancements in  multi-agent frameworks across various research fields \cite{tang2023medagents,kim2024mdagents,li2024mmedagent,hong2024argmed}, and adopting it to tackle the novel challenges we identified for MRG in the health domain. Our framework leverages specialized agents, each focusing on distinct disease categories, enabling more precise and comprehensive report generation. A specialized agent can efficiently handle specific tasks, ensuring targeted analysis and improving the accuracy of disease detection. 



Our key contributions are as follows: \\
\begin{itemize}
    \item We present the first multi-agent LLM-based framework tailored for MRG, where multiple specialized agents collaborate to generate more clinically meaningful reports.
    \item  We introduce a sentence-level task decomposition strategy to segment reports into disease-specific components, enabling fine-tuning of multiple specialized agents on different medical conditions. Additionally, we curate disease-specific subsets from the IU X-ray and MIMIC-CXR datasets, demonstrating that MRGAgents improves performance across most disease categories. 
    \item Experimental results on two widely used medical report generation benchmark datasets demonstrate superior performance compared to existing Med-LVLM baselines.
\end{itemize}

\section{Method}

\subsection{Overall Framework of MRGAgents }

\begin{figure}
    \centering

    \includegraphics[width=0.95\linewidth]{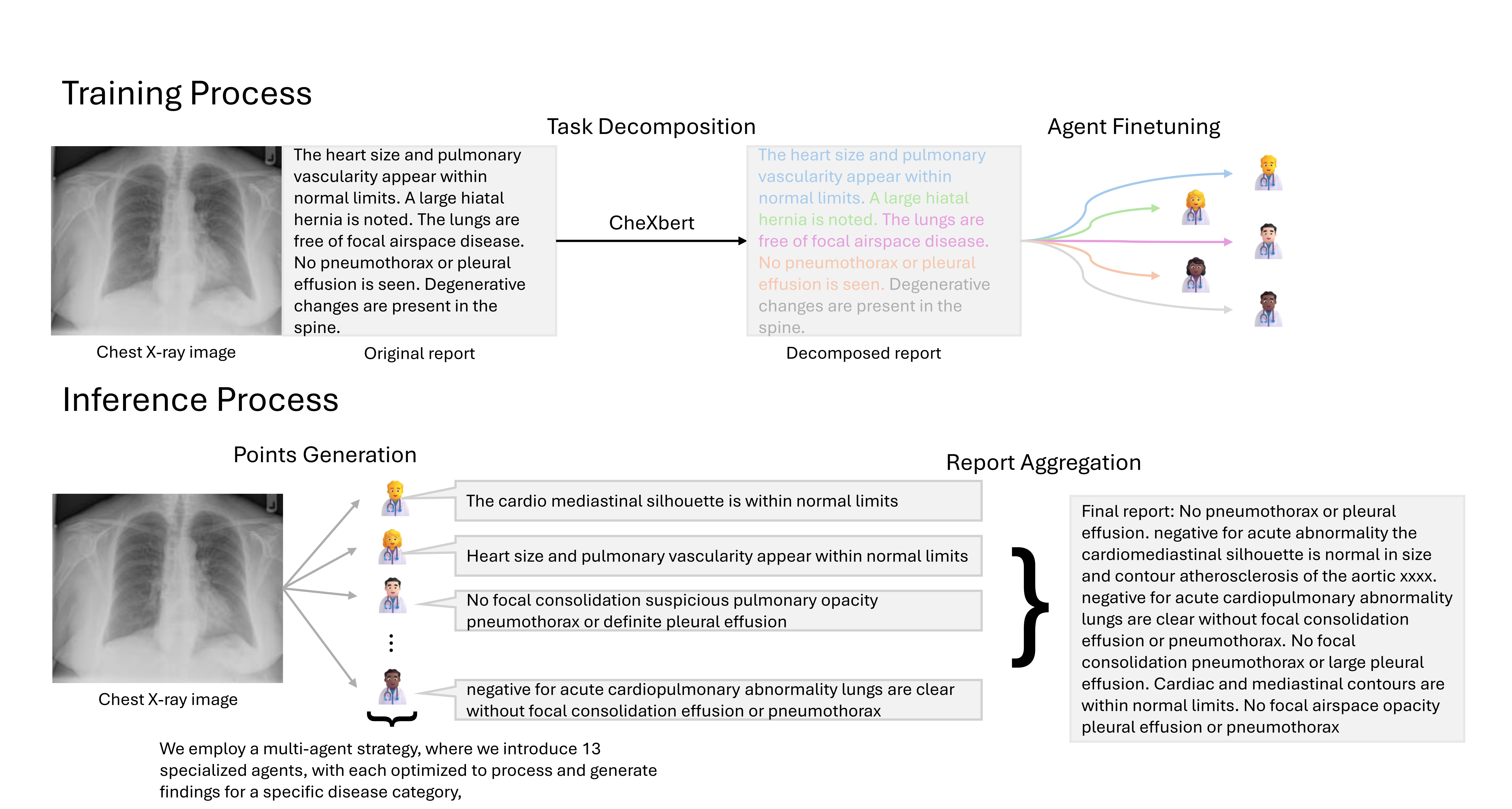}
    \caption{Overall Framework of MRGAgents}
    \label{fig:2}
\end{figure}

The MRGAgents framework, illustrated in Figure \ref{fig:2}, consists of two main steps: task decomposition and specialized fine-tuning. Instead of relying on a single Med-LVLM, it decomposes report generation into 13 disease-specific subtasks, with each fine-tuned agent generating its own report sentences. These agents correspond to the classification scheme used in medical text annotation tools like CheXbert, which categorizes reports into 14 labels, "no finding" category is exclusive. By assigning each agent to a specific category, the framework enhances precision and comprehensiveness. The generated sentences are then aggregated into a structured report.

\subsection{Task Decomposition: Sentence-Level Report Splitting }

Medical reports usually contain multiple sentences, each corresponding to different clinical information, such as normal description, abnormal finding, lesion area, etc. The generated report of existing study on medical report generation is overgeneralized, Therefore, we use sentence-level task splitting to enhance the detail and comprehensivness of the reports. 

We utilize CheXbert \cite{smit2020chexbert}, a BERT-based classifier designed for multi-label medical text classification. CheXbert categorizes radiology reports into 14 distinct observations.
It effectively identifies and labels each observation as positive, negative, or uncertain. This classifier allows us to construct subsets of training data tailored for different agent specializations. By split reports into distinct disease-related components, we ensure that each agent learns to generate precise and clinically relevant descriptions, ultimately leading to a more comprehensive and informative final report. 

\subsection{Agent Fine-tunning and Inference Process }

We fine-tune the agents in our multi-agent framework using the original training hyperparameters of BioMedGPT \cite{zhang2023biomedgpt}. During inference, each agent generates a finding sentence based on the given image. However, the generated sentences may exhibit semantic redundancy. To enhance diversity and clinical relevance, we compute each sentence’s average CIDEr score against all others, selecting the six most unique sentences with the lowest scores. This approach ensures a balance between comprehensiveness and redundancy reduction in the final report. 


\section{Experiments }

\subsection{datasets}
\begin{table}
    \centering
\caption{Sentence distribution of IU X-ray and MIMIC-CXR}
\label{tab:1}
    \begin{tabular}{lcccccc} 
        \toprule
         &  \multicolumn{3}{c}{IU X-ray}& \multicolumn{3}{c}{MIMIC-CXR}\\
 & training& validation& test& training& validation&test\\ 
        \midrule
         Enlarged Cardiomed.&  1546&  244& 452& 156,405& 1213&1394\\ 
         Cardiomegaly&  2486&  378& 720& 136,645& 1093&2107\\ 
 Lung Opacity& 2660& 362& 762& 102,548& 752&2306\\
 Lung Lesion& 310& 50& 100& 14,702& 162&313\\
 Edema& 1704& 228& 460& 75,091& 587&1367\\
 Consolidation& 242& 32& 70& 109,220& 879&1260\\
 Pneumonia& 1396& 222& 416& 32,083& 248&613\\
 Atelectasis& 162& 26& 50& 61,139& 441&1074\\
 Pneumothorax& 176& 20& 38& 198,670& 1557&2441\\
 Pleural Effusion& 3044& 436& 860& 232,670& 1850&3296\\
 Pleural Other& 3180& 456& 904& 5,635& 47&167\\
 Fracture& 38& 4& 8& 20,810& 131&324\\
 Support Devices& 202& 28& 50& 103,934& 786&2507\\
 No Finding& 230& 38& 68& 929,135& 7168&12,149\\ 
        \bottomrule
    \end{tabular}

\end{table}
We evaluated our framework on two benchmark medical report generation datasets: IU X-ray \cite{demner2016preparing} and MIMIC-CXR \cite{johnson2019mimic}. IU X-ray includes 7,470 chest X-ray images paired with 3,955 reports, while MIMIC-CXR contains 377,110 images and 277,835 reports. Both datasets follow their official splits, with IU X-ray divided into 4,118 training, 588 validation, and 2,764 testing samples, and MIMIC-CXR split into 270,790 training, 2,130 validation, and 3,858 testing samples. 

\begin{figure}[htbp]
    \centering
    \includegraphics[width=0.75\textwidth]{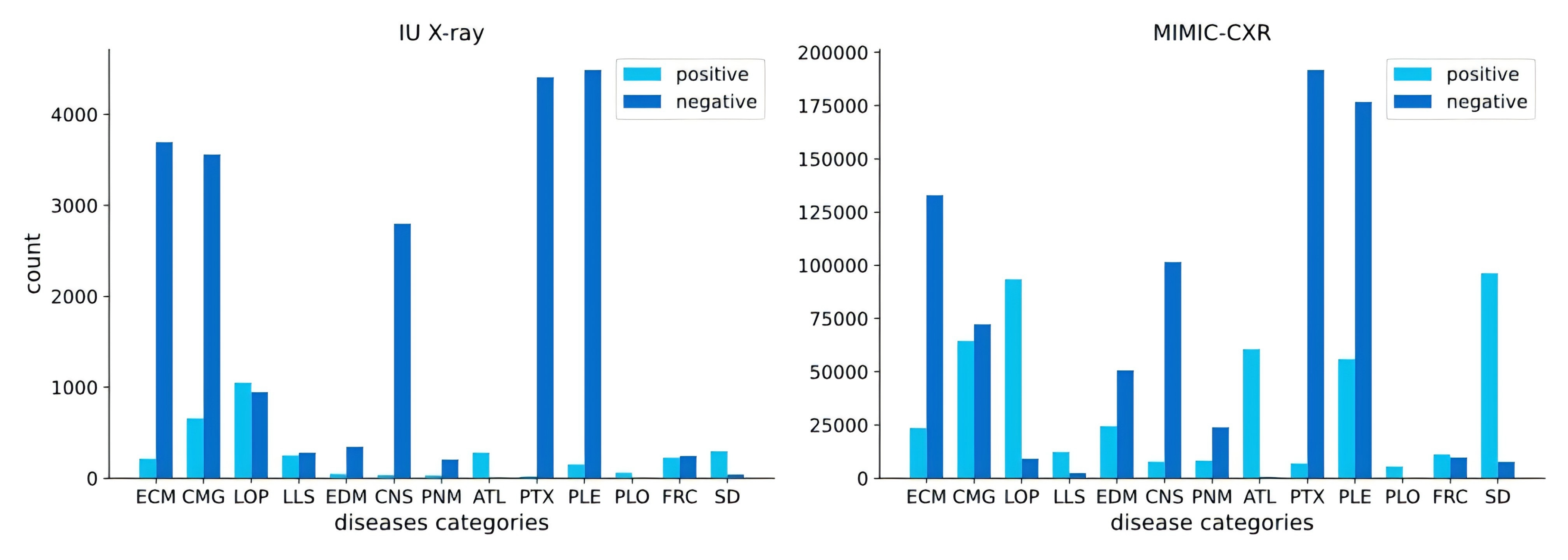}
    \caption{The distribution of positive and negative sentences in each disease.}
    \label{fig:1}
\end{figure}

We split all reports into sentences and label each sentence with CheXbert \cite{smit2020chexbert}, selecting those with positive and negative findings for training specialized agents. The sentence-level label distributions for both datasets are summarized in Table \ref{tab:1}.
In addition, in order to show the distribution of the categories more intuitively, we drew a bar chart (as shown in Figure \ref{fig:1}), with the horizontal axis representing the categories and the vertical axis representing the number of sentences.\\

\subsection{Experimental Design}

The performance of the proposed MRGAgents framework was benchmarked against established models for medical report generation to demonstrate its effectiveness.  We use BioMedGPT\footnote{Our framekwork is flexible and BioMedGPT can be replaced with any Med-LVLMs in future } as the base model for MRGAgents framework. Each agent is trained on sentences corresponding to a specific disease category, In particular, we do not train agents for 'no finding' All agents are fine-tuned based on the default setting of the fine-tuning script of BioMedGPT. Additionally, we computed accuracy and recall per disease to assess MRGAgents’ ability to correctly identify medical conditions.


\subsection{Evaluation metrics}

We evaluate MRGAgents using both natural language generation (NLG) including ROUGE-L, METEOR, and CIDEr and clinical efficacy (CE~\cite{chen-acl-2021-r2gencmn,yang2022knowledge,yang2023radiology}) including Recall, Precision and F1-score metrics.
It is important to note that the NLG metrics for MRGAgents are calculated under the assumption that the sentences describing the relevant diseases are known in advance, enabling the model to construct the final report from these pre-identified sentences. 

\section{Result and Discussion }

\subsection{Comparison with Existing Models}

Table \ref{tab:2} presents the comparison results between our MRGAgents and the state-of-the-art MRG methods of BioMedGPT, R2Gen etc etc... . 
\begin{table}
    \centering
     \scriptsize
\caption{Comparisons of our MRGAgents with state-of-the-art MRG methods. *'indicates that the results were obtained through reproduction. The best results are highlighted in bold. Methods with additional annotations are showed in gray for reference.}
\label{tab:2}
    \begin{tabular}{clccc|ccc}
        \toprule 
         Data&Model & \multicolumn{3}{c}{CE Metrics}&\multicolumn{3}{c}{NLG Metrics} \\
  & & Precision& Recall&F1
&  METEOR 
&CIDEr& ROUGE-L
\\
        \midrule
 \textbf{IU X-ray}& R2GenCMN \cite{chen-acl-2021-r2gencmn} & -& -&-
&  0.191 
&-& 0.375
\\
 & R2Gen \cite{chen2020generating} & & &
&  0.142 
&& 0.371
\\
 & LKAM \cite{yang2023radiology} & -& -&-
&  - 
&0.407& \textbf{0.399}
\\
 & BioMedGPT \cite{zhang2023biomedgpt} & 0.360& 0.354&\textbf{0.355}
&  - 
&0.401& -
\\
 & MGSK \cite{yang2022knowledge} & \textcolor[gray]{0.7}{-}& \textcolor[gray]{0.7}{-}&\textcolor[gray]{0.7}{-}&  \textcolor[gray]{0.7}{-}&\textcolor[gray]{0.7}{0.382}& \textcolor[gray]{0.7}{0.381}\\
 & \textbf{MRGAgents (ours)} & \textbf{0.369}& \textbf{0.376}&0.346
&  \textbf{0.205} 
&\textbf{0.426}& 0.331
\\
         \midrule
        \textbf{MIMIC-CXR}&R2GenCMN \cite{chen-acl-2021-r2gencmn} & 0.334& 0.275&0.278
&   0.142 
&-&  \textbf{0.278}
\\
 & R2Gen \cite{chen2020generating} & 0.333& 0.273&0.276
&  0.142 
&& 0.277
\\
 & LKAM \cite{yang2023radiology} & \textbf{0.420}& 0.339&\textbf{0.352}
&  - 
&\textbf{0.111}& 0.274
\\
 & BioMedGPT* \cite{zhang2023biomedgpt} & 0.290& 0.314&0.286
&  0.072 
&0.02& 0.144
\\
 & MGSK \cite{yang2022knowledge} & \textcolor[gray]{0.7}{0.458}& \textcolor[gray]{0.7}{0.348}&\textcolor[gray]{0.7}{0.371}&  \textcolor[gray]{0.7}{-}&\textcolor[gray]{0.7}{0.203}& \textcolor[gray]{0.7}{0.284}\\
 & \textbf{MRGAgents (ours)} & 0.369& \textbf{0.382}&0.258&  \textbf{0.144} &0.031& 0.178\\
        \bottomrule
    \end{tabular}

\end{table}
MRGAgents outperformed BioMedGPT, which is the  backbone method, on majority of NLG and CE metrics, particularly improving the CIDEr score from 0.401 to 0.426 on IU X-ray. Focusing on CE metrics, MRGAgents demonstrated consistent improvements over its baseline model. On IU X-ray, MRGAgents improves recall from 0.354 to 0.376, indicating its enhanced ability to capture clinical findings. Similarly, in MIMIC-CXR, recall increased from 0.314 to 0.382, making it the second-highest among all models, which is particularly important for reducing missed diagnoses. Precision also improved on both datasets, demonstrating that the specialized agent-based framework enhanced both sensitivity and specificity in report generation. When compared to other state-of-the-art models, MRGAgents remained highly competitive. On IU X-ray, it achieves the highest CIDEr (0.426) and METEOR (0.205) scores among all evaluated models, while maintaining strong CE metrics. In the MIMIC-CXR dataset, MGSK achieves the highest CE metrics, likely due to its external knowledge source, which supplements training data with additional clinical information. The lower ROUGE-L score may be due to the increased length and structural differences in the generated reports, which, despite being more comprehensive, reduce the longest common subsequence overlap with the reference reports. 


\subsection{Quantitative Analysis in Disease Level }

In addition to evaluating performance at the whole report level, we also assess the accuracy of disease-specific classifications. BioMedGPT is trained without task decomposition, while MRGAgents is trained using the task decomposition strategy described earlier. The results are presented in table \ref{tab:3}.
\begin{table}
      \scriptsize
    \centering
\caption{Comparison of the proposed MRGAgents with previous studies on the curated subset of IU X-ray, evaluating the accuracy and recall of disease classification. The best results are highlighted in bold. }
\label{tab:3}
    \begin{tabular}{lcccc}
        \toprule
 & \multicolumn{2}{c}{IU X-ray} & \multicolumn{2}{c}{MIMIC-CXR}\\
 &  BioMedGPT&MRGAgents & BioMedGPT&MRGAgents\\
        \midrule
         Enlarged Cardiomediastinum
&   0&\textbf{0.04}& 0.374
&\textbf{0.641}\\
         Cardiomegaly
&   \textbf{0.856}&0.850 & 0.036
&\textbf{0.758}\\
         Lung Opacity
&   0.076&\textbf{0.163}& 0.005
&\textbf{0.481}\\
         Lung Lesion
&   0&\textbf{0.480}& 0
&\textbf{0.556}\\
         Edema
&   0.035&\textbf{0.583}& 0.199
&\textbf{0.715}\\
         Consolidation
&   0.171&\textbf{0.857}& 0.818
&\textbf{0.834}\\
         Pneumonia
&   0&\textbf{0.404}& 0
&\textbf{0.261}\\
         Atelectasis
&   0&\textbf{0.160}& 0.103
&\textbf{0.987}\\
         Pneumothorax
&   0&\textbf{0.053}& 0.885
&\textbf{0.962}\\
         Pleural Effusion
&   0.288&\textbf{0.979}& 0.636
&\textbf{0.772}\\
 Pleural Other
&  0&0 & 0
&\textbf{0.904}\\
 Fracture
&  0&0 & 0
&\textbf{0.185}\\
 Support Devices&  0&\textbf{0.440}& \textbf{0.115}&0\\
        \bottomrule
    \end{tabular}

\end{table}

As shown in Table \ref{tab:3}, MRGAgents significantly improves disease-specific classification accuracy across both datasets. In IU X-ray, it enhances detection of key conditions, with consolidation increasing from 0.171 to 0.857 and Pleural Effusion from 0.288 to 0.979. It also identifies previously missed findings like Lung Lesion (0.480) and Pneumonia (0.404). In MIMIC-CXR, MRGAgents outperforms BioMedGPT in most categories, notably in Cardiomegaly (0.758 vs. 0.036), Atelectasis (0.987 vs. 0.103), and Pleural Effusion (0.772 vs. 0.636), while detecting Pleural Other (0.904) previously missed. These results demonstrate that MRGAgents enhances both the identification of critical abnormalities and disease coverage

\subsection{MRAgents Visualization}

\begin{figure}
    \centering
    \includegraphics[width=0.95\linewidth]{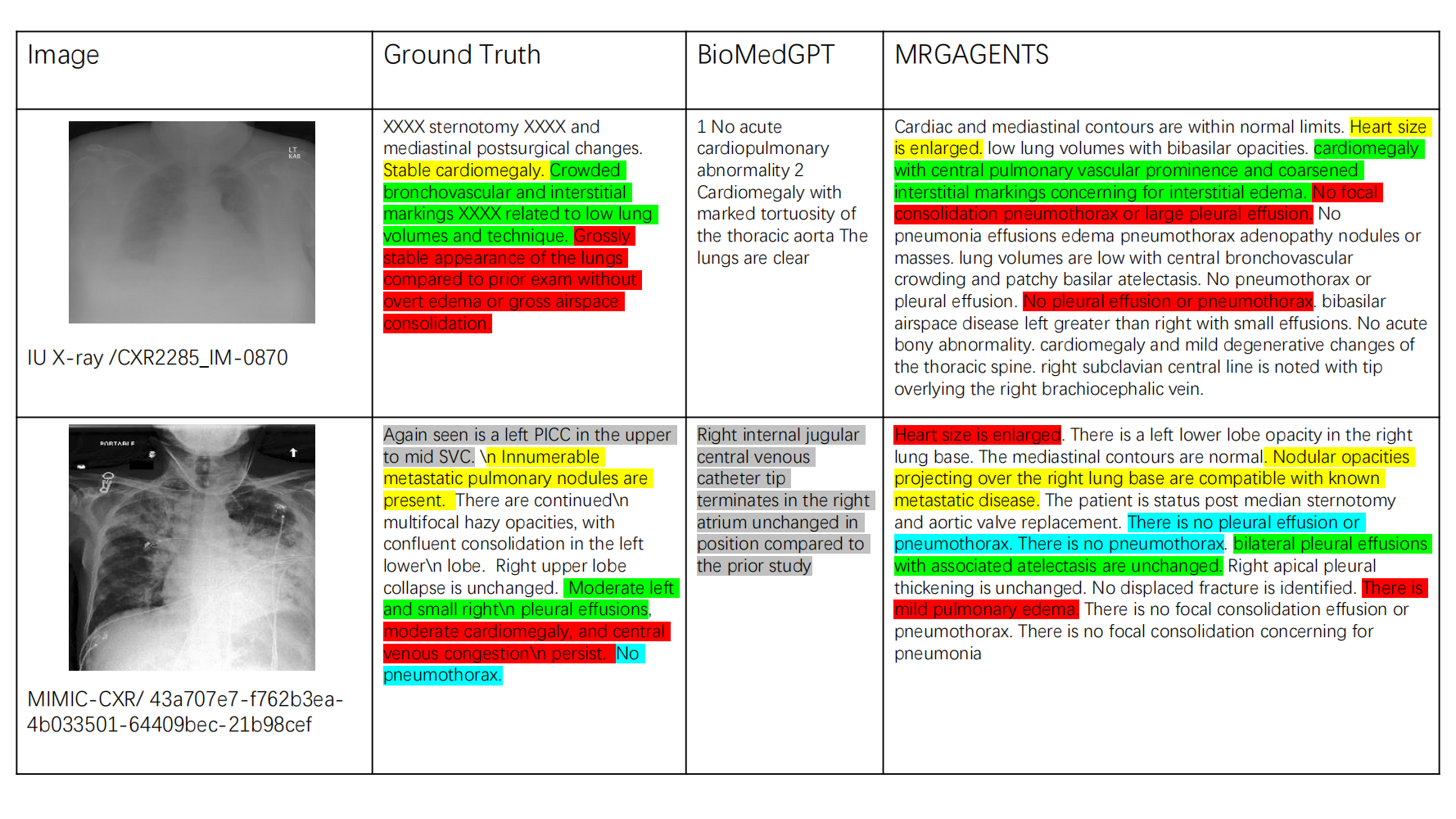}
    \caption{Examples of generated reports, with different text colors highlighting various medical descriptions for comparison with the Ground Truth. All reports generated by MRGAgents consist of 13 sentences, each corresponding to a specific disease category. }
    \label{fig:3}
\end{figure}

Figure \ref{fig:3} presents a case study comparing reports generated by MRGAgents, BioMedGPT, and the ground truth, where MRGAgents produced most comprehensive and clinically relevant reports compared to the baseline model. In the first example, MRGAgents accurately identified key clinical findings such as "cardiomegaly", "pleural effusion", and "low lung volumes", which were absent with BioMedGPT, resulting in a less informative report. This improvement stems from our agent-based approach, which systematically addresses all relevant disease categories. However, MRGAgents still has limitations, as it fails to describe support devices present in the images. This highlights the need for further refinement to improve coverage of non-disease-related findings, such as medical implants and assistive devices.

\section{Conclusion}
In this paper, we proposed MRGAgents, a novel multi-agent framework based on Med-LVLMs, designed to enhance the accuracy and clinical relevancy of automated medical report generation. By employing task-specific agents that are fine-tuned on disease-centric data, our framework ensures the generation of more detailed and clinically valuable reports. Experiment on IU X-ray and MIMIC-CXR datasets demonstrated that our MRGAgents outperformed existing single-LVLM methods.

%
%
%
\bibliography{references} 

\end{document}